\documentstyle[preprint,aps,floats,tighten]{revtex}

\begin{document}
\draft
\preprint{UTAPHY-HEP-14}
 
\title{%\vspace*{0.5in} %%%%%%%%%%%% For preprints %%%%%%%%%%%%%%%%%%%
       Relations between the SNO and the Super Kamiokande
       solar neutrino rates}
\author{Waikwok Kwong and S. P. Rosen}
\address{Department of Physics, University of Texas at Arlington,
         Arlington, Texas 76019-0059}
\date{February 12, 1996}
\maketitle

\begin{abstract}
By comparing the neutrino spectra measured by SNO and Super Kamiokande, we
obtain inequalities between the ratios of observed rate to SSM rate for the
two experiments. These inequalities apply to a possibly energy-dependent
reduction of the SSM flux and to the case of neutrino oscillations. We use
them to examine the relationship between the two experiments expected for the
MSW and ``Just-So" oscillation scenarios.
\end{abstract}

\narrowtext

% INTRODUCTION

Two high statistics solar neutrino experiments will be coming on line in the
near future. Super Kamiokande (SK)\cite{SuperK} expects to begin taking data
in April of 1996, and the Sudbury Neutrino Observatory (SNO)\cite{SNO} in
November 1996. Both experiments will observe only the $^8$B neutrinos, and
they expect about ten events a day instead of the one event seen every few
days in current experiments. Thus they will provide accurate determinations of
the solar neutrino interaction rates and of the spectral shape of the final
state  electrons\cite{spectral-ratio}. 

Here we will use a method we devised earlier\cite{semi-empirical} to compare
the neutrino spectra that are actually being measured by the two experiments,
and to derive relations between their total rates.

% PRELIMINARIES

Super Kamiokande can detect all three flavors of neutrinos through elastic
scattering with atomic electrons, $\nu\,e \to \nu\,e$. In principle it is
sensitive to neutrinos of the lowest energy, but in practice it is limited
because of backgrounds from natural radioactivity. The correlation between
the electron and the neutrino energy is poor because high energy neutrinos
can produce soft scattered electrons.

One of the principle reactions at SNO is the charged-current process $\nu_e\,d
\to p\,p\,e$, which is sensitive only to $\nu_e$. The correlation between the
electron and neutrino energy is much better than that of the elastic
scattering---since the two-proton system is relatively heavy, the electron
tends to carry off most of the neutrino energy. However, this reaction has a
threshold of 1.442 MeV, and so it is not sensitive to very  low energy
neutrinos. 
% Because Super kamiokande will detect a larger fraction of the
% spectrum of $^8$B neutrinos than SNO, it is only appropiate to ask how much
% of the spectrum measured by SNO is contained in that measured by Super
% Kamiokande.

% METHOD & RESULTS

A general expression for total rates can be written in terms of the $^8$B flux
$\phi(E_{\nu})$ from the standard solar model (SSM), an electron-neutrino
``survival probability'' $P(E_\nu)$, and an experimental cross section
$\sigma$ as
  \begin{equation}
  R = \int P(E_\nu)\,\phi(E_{\nu})\,\sigma(E_{\nu})\,dE_\nu~.
  \end{equation}
The function $P(E_\nu)$ parameterizes any, possibly energy-dependent,
differences between the SSM flux and the one that is actually measured on
Earth. These include overall reduction of neutrino fluxes due to solar
physics and energy-dependent loss of flux due to oscillations into sterile
neutrinos. All experimental parameters are hidden in the cross
section $\sigma$ which involves a convolution over an energy resolution
function, a detection efficiency, and the theoretical cross section. The
electron energy resolution for SNO is rather close to that of Super
Kamiokande, $\Delta E/E$ at 10 MeV is about 10--12\%. The detection efficiency
above trigger threshold is very close to 100\% for both experiments.
In this analysis we will use the same parameters for both experiments: 11\%
for the energy resolution and a perfect efficiency with a 5 MeV trigger
threshold. For the $\nu_e d$ theoretical cross section, we use the result of
Ref.~\cite{Kelly-Uberall}.

Since the functions $\phi\sigma(E_\nu)$ are known quantities in both
experiments, we compare their shapes by defining
  \begin{eqnarray}
  f_{\rm SK}(E_\nu) &\equiv& \frac{\phi\sigma(\nu_e e,E_{\nu})}
                  {\int\phi\sigma(\nu_e e,E_{\nu})\,dE_\nu}~,\nonumber\\[1ex]
  f_{\rm SNO}(E_\nu) &\equiv& \frac{\phi\sigma(\hbox{SNO},E_{\nu})}
                      {\int\phi\sigma(\hbox{SNO},E_{\nu})\,dE_\nu}~,
  \label{eq:shape}
  \end{eqnarray}
which are plotted in Fig.~1. Now, let us write
  \begin{equation}
  f_{\rm SK}(E_\nu) = \alpha\,f_{\rm SNO}(E_\nu) + r(E_\nu)~,
  \label{eq:compare}
  \end{equation}
and maximize the constant $\alpha$ subject to the condition that the remainder
function $r(E_\nu)$ be everywhere positive. The value obtained, $\alpha$ =
0.57, is mainly controlled by the behavior of the cross sections at the upper
end of the $^8$B spectrum: the cross section for elastic $\nu e$ scattering
rises linearly with the neutrino energy, but that for the charged-current
interaction at SNO rises much more quickly. A consequence of this behavior is
that variations at the low energy end, such as changes in the trigger
thresholds and efficiencies, have no effect on $\alpha$ to first order; they
affect $\alpha$ only indirectly through a small change in the normalization
of $\phi\sigma$. This can be seen in Table 1, where we have listed the values
of $\alpha$ for different energy resolutions for the two experiments.

\begin{table}
\caption{Dependence of $\alpha$ on the energy resolution of SNO and
         Super Kamiokande.}
{\small
$$\begin{array}{c | c | c c c |}
\multicolumn{2}{c}{} &\multicolumn{3}{c}{\hbox{SNO resolution}}\\[1ex]
\cline{3-5}
\multicolumn{2}{c|}{}    & 0.10  &  0.11 & 0.12 
\rule[1ex]{0ex}{1.5ex}\rule[-1ex]{0ex}{1.5ex} \\ \cline{2-5}
\hbox{SK}         &~0.10~&~0.569 & 0.575 & 0.581~\rule[1ex]{0ex}{1.5ex}\\ 
\hbox{resolution}~&~0.11~&~0.566 & 0.572 & 0.578~\\ 
                  &~0.12~&~0.563 & 0.569 & 0.575~\rule[-1ex]{0ex}{1.5ex}\\
\cline{2-5}
\end{array}
$$}
\end{table}

Now, we drop the term $r(E_\nu)$, multiply both sides of (\ref{eq:compare})
with $P(E_\nu)$, and integrate over $E_\nu$. This gives us an inequality
between the total rates of the two experiments. Recognizing the denominators
in (\ref{eq:shape}) to be the respective SSM rates, we express the inequality
in terms of the ratios of observed to SSM event rates for either
oscillations of solar $\nu_e$ into sterile neutrinos, or for an energy
dependent reduction of the solar $\nu_e$ flux. We define for Super Kamiokande
and SNO
  \begin{equation}
  y \equiv \frac{R(\hbox{SK})}{R_{\rm SSM}(\hbox{SK})};\qquad
  x \equiv \frac{R(\hbox{SNO})}{R_{\rm SSM}(\hbox{SNO})}~,
  \end{equation}
and obtain our first inequality:
\newlength{\eqwidth}
\setlength{\eqwidth}{\hsize}
\addtolength{\eqwidth}{-\parindent}
\addtolength{\eqwidth}{-8ex}
  \begin{equation}
  \lefteqn{\rm (I)}\makebox[\eqwidth]{$y \ge \alpha\,x$}
 	\label{eq:I}
  \end{equation}

Next, let us consider the case of oscillations of $\nu_e$ into an active
neutrino, i.e., $\nu_\mu$ or $\nu_\tau$.  The rate for SNO remains unchanged,
but that for Super Kamiokande must be modified by the additional
neutral-current scattering contributions coming from $\nu_\mu$ and $\nu_\tau$:
  \begin{eqnarray}
  R(\hbox{SK}) &=& \int\Big( P\,\phi\sigma(\nu_e e,E_\nu) +
  (1-P)\,\phi\sigma(\nu_\mu e,E_\nu) \Big) dE_\nu~, \nonumber\\
  &=& \int [0.85\,P(E_\nu) + 0.15]\,\phi\sigma(\nu_e e,E_\nu)dE_\nu~,
  \end{eqnarray}
where, $\sigma(\nu_\mu e,E_\nu)$ is the common cross section for $\nu_\mu e$
and $\nu_\tau e$ scattering. To obtain the second line, we have made the
substitution $\sigma(\nu_\mu e,E_\nu) = 0.15\,\sigma(\nu_e e,E_\nu)$,
which is a very good approximation in the energy range under
consideration\cite{approx}. It allows us to write the ratio of the actual
Super Kamiokande rate to the SSM prediction in the general form
  \begin{eqnarray}
  y &=& (1-\beta) \int P(E_\nu)\,f_{\rm SK}(E_\nu)\,dE_\nu + \beta \\[1ex]
  \beta &=& \left\{\matrix{0,&\hbox{oscillation into sterile neutrinos}\cr
        0.15, & \hbox{oscillation into active neutrinos}\cr}\right.
  \end{eqnarray}
Making use of Eq.~(\ref{eq:compare}) we find the general inequality
  \begin{eqnarray}
  y \ge (1-\beta)\,\alpha\,x + \beta
  \end{eqnarray}
which includes Eq.~(\ref{eq:I}) when $\beta=0$, and gives us our second
inequality
  \begin{equation}
  \lefteqn{\rm (II)}
  \makebox[\eqwidth]{$y \ge 0.85\,\alpha\,x + 0.15$}
  \end{equation}                                                         %(3)
for oscillations into active neutrino species when $\beta = 0.15$.

The inequalities (I) and (II) are represented graphically in Fig.~2 with the
ratio $y$ as ordinate and the ratio $x$ as abscissa. Combinations of
observations from the two experiments can be represented by points in the
diagram; when experimental errors are taken into account, the points become
regions.

Inequality (I) requires that all observed regions lie above the line $y =
\alpha\,x, ~(\alpha=0.57)$. Since this inequality has been derived under
general conditions with few assumptions regarding solar or neutrino physics,
all points below the line are unphysical. Put another way, experimental
observations falling below the line would imply a fundamental error in
present theories of the sun and solar neutrinos.

Inequality (II) defines a region above the line $y = 0.85\,\alpha\,x + 0.15,
~(\alpha=0.57)$. which is displaced vertically above $y = \alpha\,x$ by 0.15
and has a 15\% smaller slope. All points above this line are consistent with
all solutions to the solar neutrino problem, solar physics and oscillations
into active or sterile neutrinos. Points lying between the two lines are
consistent with solar physics and oscillations into sterile neutrinos;
therefore should the results from Super Kamiokande and SNO fall within this
region, we will be able to rule out oscillations into active neutrinos and
predict a smaller neutral-current signal in SNO than expected in the SSM.

We can represent the present measurements from Kamiokande II, namely, $y =
0.51 \pm 0.07$ as a horizontal band in the diagram.  Within statistical
fluctuations, the observations from Super Kamiokande are expected to fall
inside this band. 

% THE DIFFERENT APPROXIMATIONS

There are various fits
\cite{GALLEX-fit,Barger-etal,Hata-Langacker,Krastev-Petcov,Bahcall-Krastev}
to the existing solar neutrino data based upon the MSW mechanism and the
Just-So oscillations, and it is useful to see how they are represented in
our plot. The ``small angle" MSW solution can be characterized by an electron
survival probability
  \begin{equation}
  P(E_\nu) = e^{-C/E_\nu}, \label{eq:P-small}
  \end{equation}
where the constant $C$ is proportional to the product of $\sin^2 2\theta$
times $\Delta m^2$ and is close to 10 MeV in magnitude. In the standard
$\Delta m^2$--$\sin^2 2\theta$ oscillation parameter space, the allowed
small-angle region can be represented in a log-log plot of
constant-probability (or constant-rate) contours by a series of parallel
lines each corresponding to a different value for the product $\Delta m^2
\sin^2 2\theta$; in our Super Kamiokande vs SNO rate plot, each of these
lines maps into a single point in the $x$-$y$ rate-space, which represents a
specific rate for each experiment. As we move from one line to another in the
parameter space, the single points in rate-space map out a line.

To determine the equation for this line, we consider small changes in the
parameter $C$ around the value $C_0$ = 10 MeV. The survival probability can
then be written
  \begin{equation}
  P(E_\nu) = e^{-(C_0 + \Delta C)/E_\nu} 
           \approx (1 - \Delta C/E_\nu) e^{-C_0/E_\nu}, \label{eq:P-linear}
  \label{eq:expansion}
  \end{equation}
where $\Delta C$ is assumed to be much smaller than $E_\nu$. Both $y$ and $x$
are now linear in $\Delta C$, which can be eliminated to give a straight line
  \begin{equation}
  y = (1-\beta)\,B\,x + (1-\beta)\,A + \beta,
  \end{equation}
where $A$ and $B$ are calculable constants:
  \begin{eqnarray}
  &&B = B_{\rm SK}/B_{\rm SNO}, \quad 
    A = A_{\rm SK} - A_{\rm SNO} B, \nonumber\\
  &&A_{\rm SK,SNO} = \int e^{-C_0/E_\nu} 
                     f_{\rm SK,SNO}(E_\nu)\,dE_\nu\nonumber\\
  &&B_{\rm SK,SNO} = \int\hbox{$\frac{1}{E_\nu}$}e^{-C_0/E_\nu}
                     f_{\rm SK,SNO}(E_\nu)\,dE_\nu~.
  \end{eqnarray}
The appropriate lines evaluated using (\ref{eq:P-small}) instead of its linear
approximation (\ref{eq:P-linear}) are shown in Fig.~2 as thin solid lines
passing through the point (1,1), as required. The upper line ($\beta = 0.15$)
is for oscillation in to active neutrinos and the lower line ($\beta = 0$) is
for oscillation in to sterile neutrinos. These two lines are only very
slightly curved, indicating that the approximation (\ref{eq:P-linear}) is
valid for a wide range of values for $C$.

The ``large angle" MSW solution has an electron-neutrino survival probability
  \begin{equation}
  P(E_\nu) = \sin^2\theta
  \end{equation}
which is independent of energy and $\Delta m^2$. Thus it maps vertical lines
in parameter space into single points in rate-space, and as we move from one
line to another the points in rate-space trace a line. Using the above
survival probability in the expression for $x$ and $y$, we obtain the
equation of the line as
  \begin{equation}
  y = (1-\beta)x + \beta.
  \end{equation}
It is a straight line that always passes through the point (1,1) corresponding
to no oscillations, and becomes $y = x$ in the sterile case ($\beta$ = 0). It
is plotted in Fig.~2 as the dot-dashed lines for the active and sterile cases.

In the Just-So solution, the electron neutrino survival probability is given
by
  \begin{equation}
  P(E_\nu) = 1-\sin^2 2\theta\,\sin^2\left(\frac{\Delta m^2 L}{4E_\nu}\right).
  \end{equation}
The value of $\Delta m^2$ must be chosen to yield an oscillation length
of the same order as the Earth--Sun distance $L$. Thus, for some energy $E_0$
within the spectrum of solar neutrinos
  \begin{equation}
  \frac{\Delta m^2 L}{4 E_0} = (n + \hbox{$\frac{1}{2}$}) \pi~.
  \end{equation}
Letting $\Delta m^2 = A\times 10^{-11} {\rm~eV}^2$ and measuring $E_\nu$ in
MeV, we can express the $y$ and $x$ coordinates as 
  \begin{eqnarray}
  \label{eq:justso-SK}
  y &=& 1 - (1-\beta)\sin^2 2\theta \int
        \sin^2\left(\frac{1.90 A}{E_\nu}\right)f_{\rm SK}dE_\nu\\
  x &=& 1 - \sin^2 2\theta \int
        \sin^2\left(\frac{1.90 A}{E_\nu}\right)f_{\rm SNO}dE_\nu
  \label{eq:justso-SNO}
  \end{eqnarray}
For specific values of $\Delta m^2$, or $A$, the two integrals can be
integrated numerically. Again, eliminating $\sin^2 2\theta$ from the two
equations gives us a linear relationship between $x$ and $y$. As $\sin^2
2\theta$ is varied from 0 to 1, the point $(x,y)$ traces a straight line
starting from $(1,1)$ and ends at a point $(x_0,y_0)$ with $x_0 >0$ and
dependent on the value of $\Delta m^2$. By varying also $\Delta m^2$, the
entire parameter space is mapped into finite regions in Fig.~3: oscillations
into active neutrinos give rise to the area enclosed by the solid curve and
oscillations into sterile neutrinos give rise to the one enclosed by the
dotted curve. A point falling outside these two regions cannot be explained
using the Just-So oscillations.

So far, these lines and closed regions we have discussed represent the entire
parameter space within the individual approximations. Existing data from
Kamiokande II, the Chlorine experiment, and the two gallium experiments
GALLEX and SAGE favor certain ranges of the oscillation parameters. For this
we use the global fit of Ref.~\cite{Hata-Langacker} for the small- and
large-angle MSW solutions (the large-angle solution for sterile neutrinos has
been ruled out according to this fit) and the result of
Ref.~\cite{Krastev-Petcov} for the Just-So solution (depending on the how the
fitting is done, the sterile case can also be rule out here,
see~\cite{Krastev-Petcov} for details). Both analyses took into consideration
theoretical uncertainties. The allowed regions at 95\% confidence
from these constraints on the SNO and Super Kamiokande rates are shown in
both Fig.~2 and 3 as heavy black lines and shaded patches.

% DISCUSSIONS

By considering the overlap of the regions corresponding to different solar
neutrino solutions, we can anticipate the implications of measurements to be
made by SNO and Super Kamiokande. 

Our first observation is that the Just-So solutions occupy the largest area
in the rate-space of the two experiments and are therefore the most difficult
ones to rule out. From the total rates of SNO and Super Kamiokande alone, it
would be practically impossible to rule out the Just-So oscillations without
also ruling out both the small- and large-angle MSW solutions. There is only a
very small window with $x\lesssim0.16$ and $0.15\le y \lesssim 0.25$ for
active neutrinos, or $0\le y \lesssim 0.15$ for sterile neutrinos, in which
this is possible. In contrast, the large- and small-angle MSW solutions
occupy zero area in rate-space. This makes them extremely sensitive to the
Super Kamiokande and SNO measurements: the data point must falls right on top
of one of the lines in Fig.~2, to within experimental uncertainty. 

Our second observation is that from the total rates alone, it would be
difficult to distinguish between the large- and small-angle MSW solutions at
the 3-$\sigma$ level. The solar neutrino rate of Super Kamiokande is about 50
times that of Kamiokande II; in about five years Super Kamiokande will have
accumulated 50 times as much data as the present Kamiokande II. This
translates into a factor of seven in the statistical uncertainty so that the
ratio of observed to SSM rate for Super Kamiokande will have a 1$\sigma$
uncertainty of $\pm0.01$ (instead of $\pm0.07$ for Kamiokande II), provided
that it is not limited by systematic uncertainties. On the other hand, Fig.~2
shows that the maximum distance in the $y$ direction between the large- and
small-angle lines are only about 0.02. It would be easier to distinguish
between the sterile and the active case of the MSW solutions, especially if
both experiments yield rates that are no larger than about half their SSM
values.

With the help from the four existing experiments, some of these difficulties
may be overcome. For example, depending on where the future data point falls,
we may be able to distinguish the large-angle MSW solution for active neutrinos
(the short heavy black line in Fig.~2 and 3) from the small-angle one but
probably not from the Just-So solution. 

Implications obtained from rate measurements can be tested by examining the
spectra of recoil electrons observed in both Super Kamiokande and SNO.
Although the differences tend to be rather subtle, the combination of high
statistics and the ``normalized spectral ratio" method\cite{spectral-ratio}
should enable us to distinguish between active and sterile neutrinos.
In addition, the Just-So solution is much more sensitive to the BOREXINO
experiment\cite{BOREXINO} than to either Super Kamiokande or SNO because of
the monenergetic $^7$Be lines. This should help us separate the Just-So
oscillations from the MSW solutions.

% ACKNOWLEDGEMENT

This work was supported in part by the U.S. Department of Energy Grant 
No.~DE-FG03-96ER40943. The authors would like to thank Gene Beier and Hank
Sobel for providing the experimental parameters for SNO and Super Kamiokande.
One of the authors (SPR) would like to thank Geoffrey West and the Los Alamos
National Laboratory for their hospitality at the 1995 Summer Workshop and
Plamen Krastev for a conversation which initiated this work.

\newpage

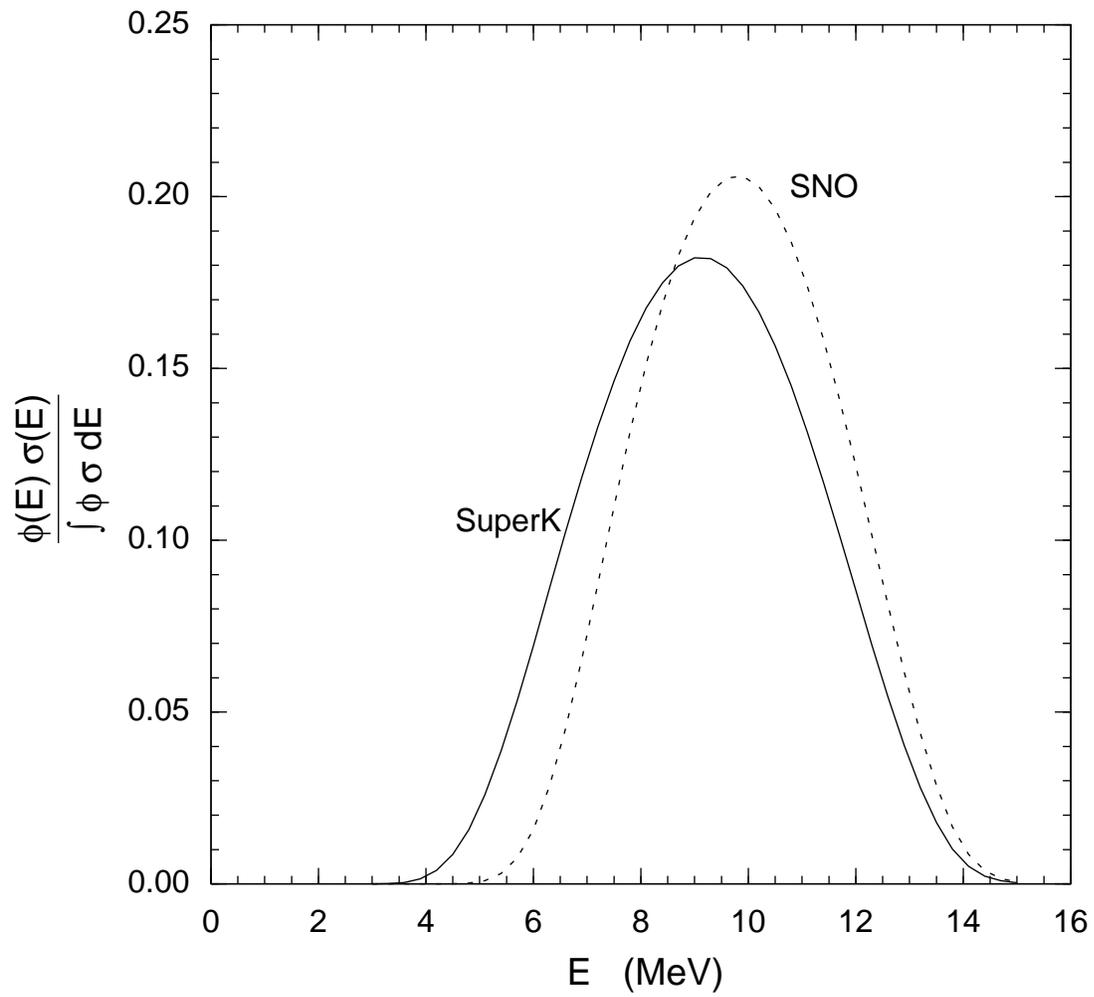
\begin{figure}
\caption{The normalized shapes of $\phi\sigma$ for SNO and Super Kamiokande.}
\end{figure}

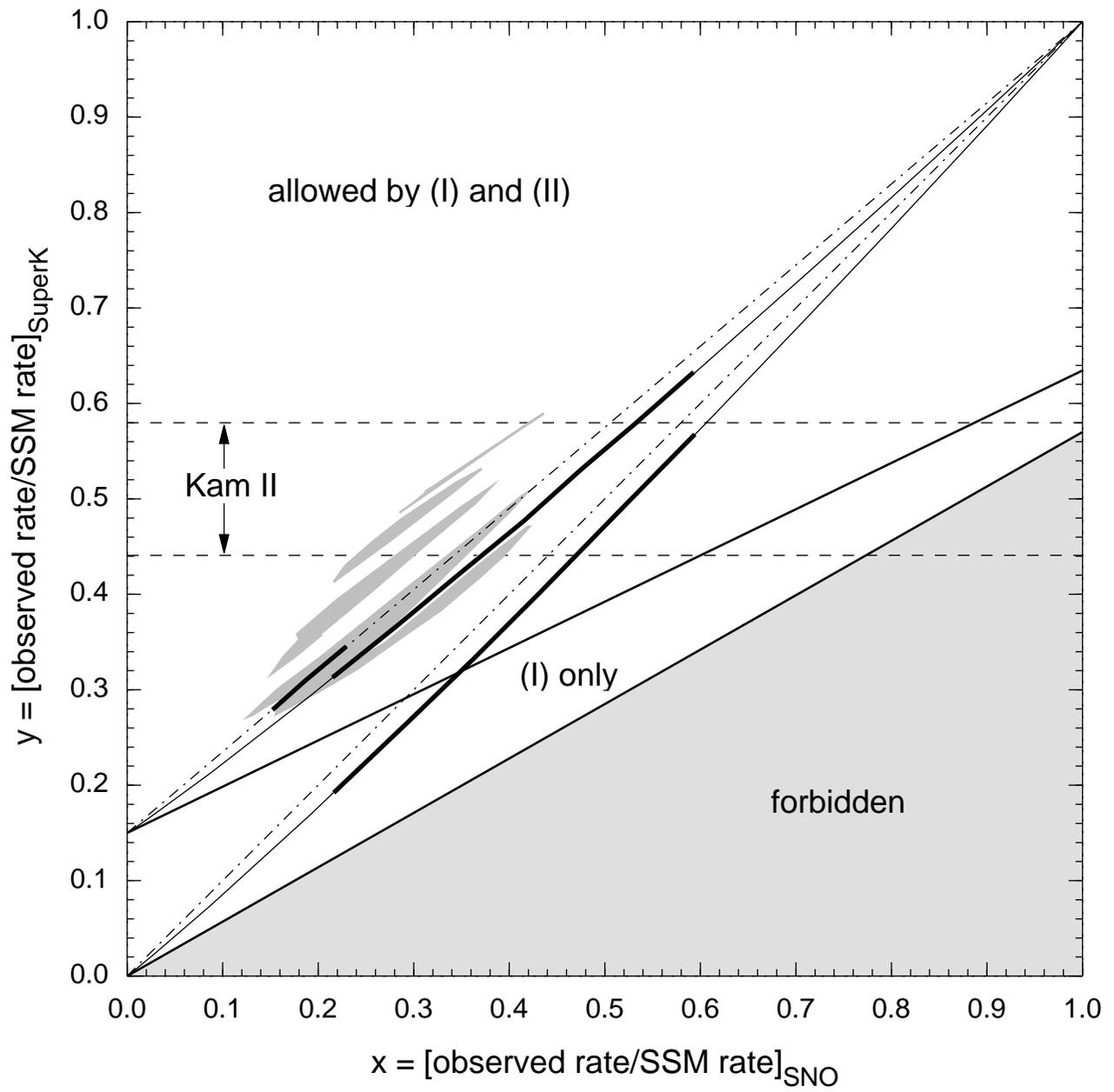
\begin{figure}
\caption{The MSW solutions in the Super Kamiokande-SNO rate-space.
The inequalities (I) and (II) divide the rate-space into three regions
labeled  ``allowed by (I) and (II)", ``(I) only", and ``forbidden". The
small-angle MSW solution must lie on the solid thin lines and the large-angle
solution must lie on the dot-dashed lines. The upper pair is for oscillation
into active neutrinos and the lower pair for sterile neutrinos. Bounds from
existing data are represented by the heavy black lines. The patches of shaded
areas are bounds from the Just-So solution, shown here for comparison.}
\end{figure}

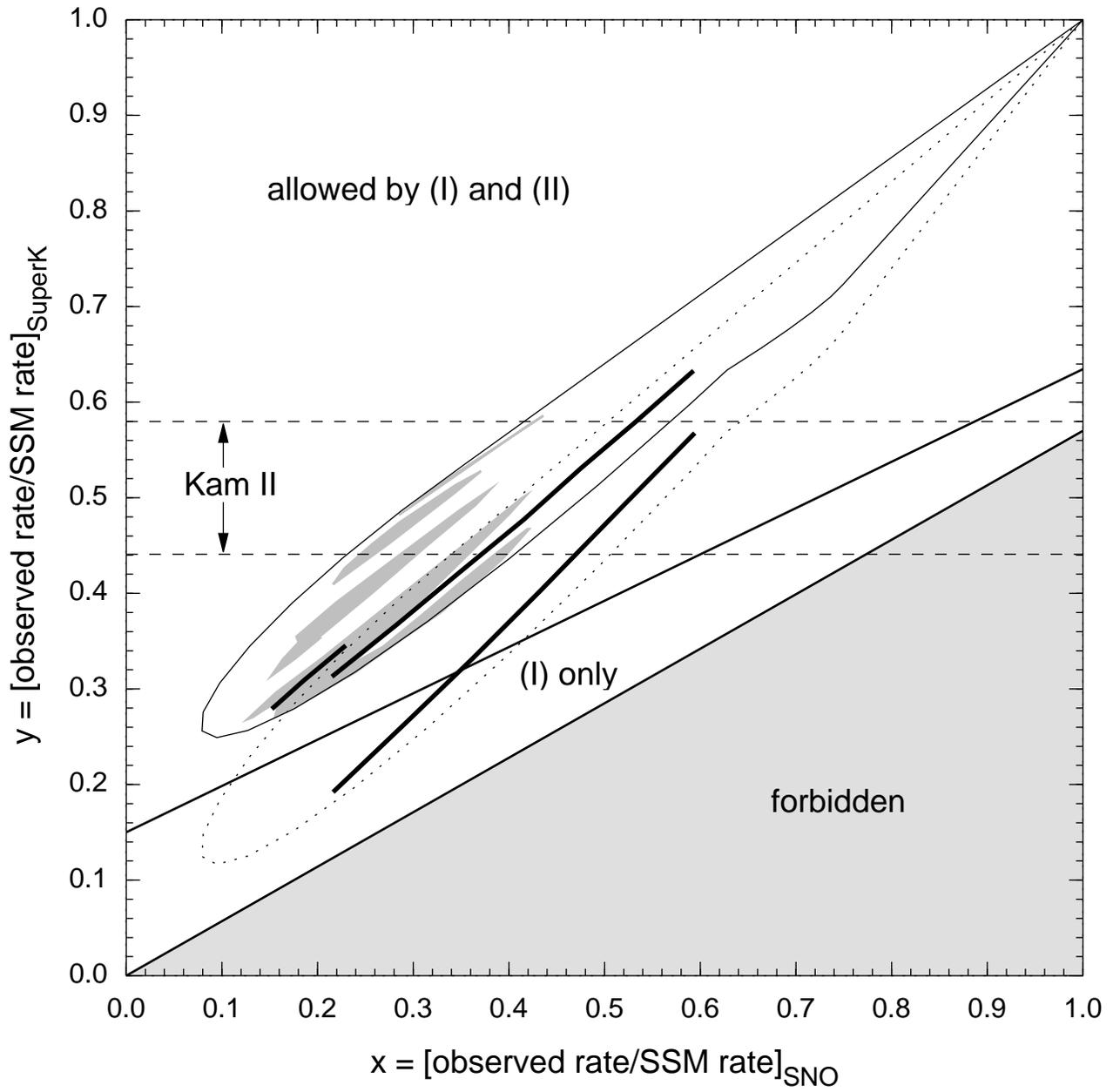
\begin{figure}
\caption{The Just-So solutions in the Super Kamiokande-SNO rate-space.
The region bounded by the thin solid and dotted curves are the solution
spaces for Just-So oscillations into active and sterile neutrinos
respectively. See also Fig.~2.} \end{figure}

\end{document}